\documentclass[natbib]{svjour3}                     % onecolumn (standard format)

\usepackage{url}
\usepackage{epsfig}
\usepackage{graphicx}
\usepackage{color}

\newcommand{\beq}{\begin{equation}}
\newcommand{\eeq}{\end{equation}}

\newcommand{\aap}{{Astron. Astrophys.}}
\newcommand{\apj}{{Astrophys. J.}}
\newcommand{\apjl}{{Astrophys. J. Lett.}}
\newcommand{\apjs}{{Astrophys. J. Suppl.}}

\newcommand{\mnras}{{Mon. Not. Roy. Astron. Soc.}}
\newcommand{\araa}{{Annual Rev. Astron. \& Astrophys.}}
\newcommand{\grl}{{Geophys. Res. Lett.}}
\newcommand{\jgr}{{J. Geophys. Res.}}
\newcommand{\nat}{{Nature}}

\newcommand{\pasj}{{Pub. Astron. Soc. Japan}}
\newcommand{\apss}{{Astrophys. Space Sci.}}
\newcommand{\ssr}{{Space Sci. Rev.}}

\begin{document}

%**************************************************************

\title{Magnetic Reconnection in Extreme Astrophysical Environments\thanks{Work supported by NSF under Grant PHY-0903851}}

\author{Dmitri A. Uzdensky}

%\affil{Center for Integrated Plasma Studies, Department of Physics, UCB-390, University of Colorado, Boulder CO 80309}
\institute{Center for Integrated Plasma Studies, Department of Physics, UCB-390, University of Colorado, Boulder CO 80309.  \email{uzdensky@colorado.edu}}

%\date{Oct. 7, 2008 - \today}
\date{Received: September 24, 2010 / Revised: December 20, 2010 / Accepted: January 4, 2011}

\maketitle

%**************************************************************

\begin{abstract}

Magnetic reconnection is a fundamental plasma physics process of breaking ideal-MHD's frozen-in constraints and of dramatic rearranging of the magnetic field topology, which often leads to a violent release of the free magnetic energy. Most of the magnetic reconnection research done to date has been motivated by the applications to systems such as the solar corona, the Earth magnetosphere, and magnetic confinement devices for thermonuclear fusion. These environments have relatively low energy densities and the plasma is adequately described as a mixture of equal numbers of electrons and ions and where the dissipated magnetic energy always stays with the plasma.
In contrast, in this paper I would like to introduce a different, new direction of research --- reconnection in high energy density radiative plasmas, in which photons play as important a role as electrons and ions; in particular, in which radiation pressure and radiative cooling become dominant factors in the pressure and energy balance. 
This research is motivated in part by rapid theoretical and experimental advances in High Energy Density Physics, and in part by several important problems in modern high-energy astrophysics. 
I first discuss some astrophysical examples of high-energy-density reconnection and then identify the key physical processes that distinguish them from traditional reconnection. Among the most important of these processes are: special-relativistic effects; 
radiative effects (radiative cooling, radiation pressure, and radiative resistivity); and, at the most extreme end --- QED effects, including pair creation. 
The most notable among the astrophysical applications are situations involving magnetar-strength fields ($10^{14} - 10^{15}$~G, exceeding the quantum critical field $B_* \simeq 4\times 10^{13}$~G). The most important examples are giant flares in soft gamma repeaters (SGRs) and magnetic models of the central engines and relativistic jets of Gamma Ray Bursts (GRBs). The magnetic energy density in these environments is so high that, when it is suddenly released, the plasma is heated to ultra-relativistic temperatures. As a result, electron-positron pairs are created in copious quantities, dressing the reconnection layer in an opticaly thick pair coat, thereby trapping the photons. The plasma pressure inside the layer is then  dominated by the combined radiation and pair pressure. At the same time, the timescale for radiation diffusion across the layer may, under some conditions, still be shorter than the global (along the layer) Alfven transit time, and hence radiative cooling starts to dominate the thermodynamics of the problem. The reconnection problem then becomes essentially a radiative transfer problem. In addition, the high pair density makes the reconnection layer highly collisional, independent of the upstream plasma density, and hence radiative resistive MHD  applies. The presence of all these processes calls for a substantial revision of our traditional physical picture of reconnection when applied to these environments and thus opens a new frontier in reconnection research.

\keywords{magnetic reconnection \and magnetars \and radiative processes \and SGR Flares \and Gamma-ray bursts}

\end{abstract}

%**************************************************************
\newpage

\section{Introduction}
\label{sec-intro}

%**************************************************************

Magnetic reconnection is an important  fundamental  plasma process involving a rapid change of magnetic topology and often leading to a violent release of magnetic energy. It is widely regarded as one of the most important and ubiquitous 
phenomena in plasma physics, with numerous applications in laboratory plasma physics, 
space and solar physics, and astrophysics~\citep{Zweibel_Yamada-2009, Yamada_etal-2010}.
In particular, reconnection is believed to be the underlying mechanism for many of the most spectacular and energetic 
phenomena in various natural and unnatural plasma environments throughout the solar system. 
The most prominent examples include the solar corona (solar flares),  the Earth magnetosphere (magnetic sub-storms and flux-transfer events), and laboratory magnetic fusion devices, such as tokamaks (sawtooth crashes). 
Furthermore, magnetic reconnection has now become a subject of dedicated laboratory plasma experiments 
(see~\cite{Zweibel_Yamada-2009,Yamada_etal-2010} for recent reviews)
%{\tt  \cite[e.g][]{Stenzel_Gekelman-1981, Anna Frank, Ono,  Yamada_etal-1997, MRX, VTF, SSX, RcX, etc, Tobin..., e.g., Ji~et~al.~1998; Brown~et~al~1999; Yamada~et~al.~2006; Egedal~et~al.~2007)}, 
and in-situ spacecraft observations in the Earth magnetosphere~\citep{Oieroset_etal-2001, Mozer_etal-2002} and in the solar wind~\citep{Phan_etal-2006}.

Because of its importance, reconnection has been the subject of many l theoretical and computational studies, especially in the last two decades.  These studies have resulted in a great progress in our understanding of reconnection, in both collisional and collisionless regimes, including the elucidation of the nature of collisionless reconnection, 
the condition for the transition from the slow resistive regime to the fast collisionless regime, the understanding of the role of the secondary tearing instability and of turbulence, and other topics. 

It is worth noting, however, that most of the reconnection research has so far been driven by our desire to understand magnetic dissipation in the above space, solar, and laboratory plasmas. Importantly, these environments are rather tenuous and hence are adequately described by traditional, low-energy-density plasma physics, i.e., as a collection of charged particles (electrons and ions)  whose numbers are conserved, and no photons.
That's why the absolute majority of traditional reconnection research and modeling was done for this type of plasmas. 

At the same time, reconnection is now also being increasingly recognized as an important physical process in numerous  astrophysical contexts beyond the solar system, especially in high-energy astrophysics~\cite[e.g.,][]{Uzdensky-2006}. 
It has been frequently invoked in systems such as:  
stellar X-ray flares~\cite[e.g.,][]{Cassak_etal-2008}; 
star-disk magnetic interaction in young stellar objects~\citep{vanBallegooijen-1994, Hayashi_etal-1996, Goodson_etal-1997, Feigelson_Montmerle-1999, Uzdensky_etal-2002a, Uzdensky_etal-2002b, Uzdensky-2004, DAngelo_Spruit-2010},  
in accreting neutron stars~\citep{Aly_Kuijpers-1990, vanBallegooijen-1994},
 and white dwarfs~\citep{Warner_Woudt-2002}; 
accretion disk coronae~\citep{GRV-1979, Haardt_Maraschi-1991, Tout_Pringle-1996, Romanova_etal-1998, DiMatteo_etal-1999, Liu_etal-2003, Schopper_etal-1998, Uzdensky_Goodman-2008, Goodman_Uzdensky-2008}; 
interstellar medium and star formation~\citep{Zweibel-1989, Lesch_Reich-1992, Brandenburg_Zweibel-1995, Lazarian_Vishniac-1999, Heitsch_Zweibel-2003};
pulsar magnetospheres and pulsar winds~\citep{Coroniti-1990, Michel-1994, Lyubarsky_Kirk-2001, Lyubarsky-2003, Kirk_Skjaeraasen-2003, Contopoulos-2007, Arons-2007, Petri_Lyubarsky-2007, Spitkovsky-2008, Lyutikov-2010}; 
black-hole jets (including AGN/Blazars)  \citep{Romanova_Lovelace-1992, Larrabee_etal-2003, Lyutikov-2003a, Jaroschek_etal-2004b, Giannios_etal-2009, Giannios_etal-2010, Nalewajko_etal-2010};
magnetar flares, especially giant flares in Soft-Gamma Repeaters (SGRs)~\citep{Thompson_Duncan-1995, Thompson_Duncan-2001, Lyutikov-2003b, Lyutikov-2006a, Uzdensky-2008, Masada_etal-2010}; and
Gamma-Ray Bursts (GRBs)~\citep{Spruit_etal-2001, Lyutikov_Blackman-2001, Lyutikov_Blandford-2002, Drenkhahn_Spruit-2002, Giannios_Spruit-2005, Giannios_Spruit-2006, Giannios_Spruit-2007, Rees_Meszaros-2005, Uzdensky_MacFadyen-2006, McKinney_Uzdensky-2010}.

Not surprisingly, when attacking these challenging problems, astrophysicists are naturally tempted to take advantage of the knowledge and physical insights obtained from the traditional, solar-system reconnection studies (including laboratory experiments) and apply them in the above astrophysical contexts.
However, it is important to appreciate that the Universe is a very big and diverse place.  I would like to emphasize that the range of possible physical conditions found in various astrophysical environments is very broad and far exceeds that encountered within our solar system. In particular, there are some astrophysical phenomena, especially in high-energy astrophysics,  where, on the one hand, reconnection has been hypothesized to play an important role, and, on the other hand, where the physical parameter regimes and hence the underlying microscopic physics are vastly different from traditional reconnection (e.g., those in solar flares, EarthÕs magnetosphere, and laboratory plasmas). Therefore, a straightforward extrapolation of conventional reconnection scalings to some of these "extreme" systems is not justified. The necessity to understand how such systems work then requires us to develop new theories of magnetic reconnection taking into account physical processes that are not usually included in conventional reconnection studies.

Some of the most important among these additional physical processes in high-energy astrophysical reconnection are those related to the large strength of the magnetic fields and hence to the overall high level of {\it energy density} in these systems. 
Among these additional "exotic" high-energy-density (HED) physical effects in high-energy astrophysics, 
are: (1) special relativity;  (2) radiation;  (3) pair creation. Note that these effects may be considered "exotic" from the point of view of a traditional reconnection researcher with a background in laboratory plasma physics or space physics, 
but they are not at all alien to a typical high-energy astrophysicist. Let me briefly discuss these effects here. 

%--------------------------------------------------------------------------

{\bf (1) Relativistic reconnection:} In some magnetically-dominated environments with {\it low} ambient plasma density, e.g., in pulsar magnetospheres and in jets of Active Galactic Nuclei (AGN), one often has to deal with {\it relativistic reconnection}. 
In particular,  {\it special-relativistic effects} are important when the reconnecting (upstream) magnetic field~$B_0$ is so strong and the plasma is so tenuous that the magnetic energy density exceeds not only the pressure, but also the rest-mass energy density $\rho_0 c^2$ of the upstream plasma. The condition for this is usually cast  in terms of the so-called  $\sigma$-parameter, $\sigma \equiv B_0^2/4\pi \rho_0 c^2 \gg 1$.
Then, the corresponding Alfv\'en speed, $V_A = cB_0/\sqrt{4 \pi \rho c^2 + B_0^2}$, and hence the reconnection outflow velocity, approach the speed of light, and one needs to understand the role of various special-relativistic effects, e.g., Lorentz contraction and time dilation~\citep{Blackman_Field-1994, Lyutikov_Uzdensky-2003, Lyubarsky-2005, Watanabe_Yokoyama-2006, McKinney_Uzdensky-2010}. 
In particular, the early efforts to build special-relativistic resistive-MHD generalizations of the Sweet--Parker and Petschek models by \cite{Blackman_Field-1994} and of the Sweet--Parker model only by \cite{Lyutikov_Uzdensky-2003} were eventually followed by the analysis by~\cite{Lyubarsky-2005}. He showed that in the Sweet--Parker case the reconnection process is slow and the reconnection inflow velocity is sub-relativistic while the reconnection outflow velocity is only mildly relativistic, due to the large inertia of the relativistically hot plasma. In contrast, in the Petchek case it was found that the strong plasma compression across the relativistic shock may lead to an ultra-relativistic outflow. However, as is always the case with the Petschek model, the actual reconnection rate could not be calculated straight-forwardly from first principles and essentially had to be assumed; only the maximum Petschek reconnection rate could be estimated and found to be essentially the same as its non-relativistic counterpart, $v_{\rm in, max} \simeq \pi/4 \ln S$, where $S\equiv Lc/\eta$ is the relativistic Lundquist number. After these analytical papers, most of the subsequent work on resistive relativistic reconnection was numerical and concentrated on Petschek's reconnection, which was achieved, by similarity with non-relativistic studies, by employing an ad hoc prescription for spatially localized anomalous resistivity~\citep{Watanabe_Yokoyama-2006, Zenitani_etal-2009}.  These studies have largely confirmed the main aspects of Lyubarsky's analytical theory of relativistic Petschek's reconnection, in particular, regarding the ultra-relativistic outflow velocity.

Relativistic reconnection is especially relevant to astrophysical {\it pair} plasmas, with one of the most important and cleanest examples being radio-pulsar magnetospheres and pulsar winds. Due to a pulsar's rapid rotation, its magnetosphere inevitably opens up and develops a large-scale equatorial current sheet beyond the light cylinder (similar to the equatorial current sheet in the solar wind). Reconnection in this current sheet may be an important factor in controlling the magnetosphere structure and hence the pulsar braking index~\citep{Arons-2007, Contopoulos-2007} and for producing the observed pulsed high-energy (x-ray and $\gamma$-ray) emission and perhaps even the coherent radio emission~\cite{Lyubarsky-1996, Arons-2007, Spitkovsky-2008, Umizaki_Shibata-2010}. At larger distances, in the pulsar wind region well beyond the light cylinder,  magnetic reconnection has been invoked to resolve the famous Òsigma problemÓ --- the expected very efficient conversion of magnetic energy to particle energy~\citep{Coroniti-1990, Lyubarsky_Kirk-2001, Kirk_Skjaeraasen-2003}. 

In addition, relativistic reconnection in pair plasmas is conjectured to be responsible for powering spectacular giant magnetar flares in Soft-Gamma Repeaters~\citep{Thompson_Duncan-1995, Thompson_Duncan-2001, Lyutikov-2003b, Lyutikov-2006a, Uzdensky-2008, Masada_etal-2010} and it may also be important to relativistic jets of GRBs~\cite{Spruit_etal-2001, Lyutikov_Blackman-2001, Lyutikov_Blandford-2002, Drenkhahn_Spruit-2002, Giannios_Spruit-2005, Giannios_Spruit-2006, Giannios_Spruit-2007, McKinney_Uzdensky-2010}  and AGNs~\citep{Lyutikov-2003a, Jaroschek_etal-2004b, Giannios_etal-2009, Giannios_etal-2010, Nalewajko_etal-2010}.

Among the key questions in theoretical studies of relativistic reconnection are: What factors control the outflow velocity and the reconnection rate in relativistic reconnection? Does reconnection fundamentally change its behavior in the relativistic regime? How is the released energy partitioned? How efficient is non-thermal particle acceleration? Is collisionless pair reconnection fast even though there is no Hall term in the corresponding generalized Ohm law?

Relativistic aspects of extreme astrophysical reconnection are probably the easiest to understand, as compared to the other two aspects discussed below (radiation and pair creation). In fact, even though our understanding of relativistic reconnection is still not mature and lags far behind that of non-relativistic reconnection, this topic has probably already grown out of its infancy stage and entered its ÒyouthÓ stage. A significant amount of work has already been done both on its fundamental physics issues \citep{Blackman_Field-1994, Lyutikov_Uzdensky-2003, Lyubarsky-2005, Jaroschek_etal-2004a, Hesse_Zenitani-2007, Zenitani_Hoshino-2008, Zenitani_etal-2009, Komissarov-2007} 
and on astrophysical applications~\citep{Coroniti-1990, Lyubarsky_Kirk-2001, Lyutikov_Blackman-2001, Lyutikov-2003a, Giannios_Spruit-2005, Giannios_Spruit-2006,  Giannios_etal-2009, Giannios_etal-2010, Nalewajko_etal-2010}. 
In particular, there has been a fair amount of theoretical~\cite[e.g.,][]{Coroniti-1990, Lyubarsky_Kirk-2001} and numerical \cite{Zenitani_Hoshino-2001, Jaroschek_etal-2004a, Zenitani_Hoshino-2005, Zenitani_Hoshino-2007, Hesse_Zenitani-2007, Zenitani_Hoshino-2008,  Zenitani_etal-2009, Liu_etal-2010} work on relativistic reconnection in {\it pair} plasmas. 
Among the most notable in this area has been a series of Particle-In-Cell (PIC) studies by Hoshino \& Zenitani \citep{Zenitani_Hoshino-2001, Zenitani_Hoshino-2005, Zenitani_Hoshino-2007, Zenitani_Hoshino-2008}.
This work has resulted in some important insights regarding a very important issue of nonthermal particle acceleration. 
Their findings can be summarized as follows. It was found that the effectiveness of nonthermal particle acceleration in relativistic pair reconnection depends strongly on the presence of a guide magnetic field. Without a strong guide field, the current sheet is violently unstable to the relativistic drift-kink instability that quickly breaks it up and broadens it and leads to strong plasma heating but no efficient particle acceleration. In contrast, a strong guide field suppresses this instability. As a result, in addition to thermal heating,  a very effective particle acceleration takes place in this case, leading to a substantial power-law tail. 
Since nonthermal relativistic  electrons (and positrons) are needed to produce observable  power-law synchrotron and very high-energy gamma-ray emission, these findings should have important consequences for interpreting observations, in particular, for restricting the magnetic field geometry~\citep{Nalewajko_etal-2010}.

I would like to note,  however, that most of the PIC simulations (both 2D and 3D) of relativistic reconnection have been done only for {\it pair} plasmas. This choice has been motivated both by the relevance of pairs to pulsar magnetospheres and winds and by the lower computational cost. In contrast, relativistic reconnection in {\it electron-ion} plasmas has not yet received as much attention and still remains an open fundamental problem in modern plasma astrophysics. 

We should realistically expect significant theoretical and computational progress in the area or relativistic reconnection in the near future, with relatively modest modifications to the existing numerical and analytical models. In fact, most of the modern PIC codes used to study non-relativistic collisionless reconnection are intrinsically relativistic and hence can be (and are being) applied to relativistic reconnection problems. In addition, recent advances in developing relativistic resistive MHD algorithms~\citep{Komissarov-2007} instill hope that more progress will be made on collisional relativistic reconnection. On the other hand, laboratory investigations of relativistic reconnection are probably not yet on the horizon; however, perhaps we should start thinking about them.

%--------------------------------------------------------------------------

{\bf (2)} {\bf Radiation:} At high energy densities various {\it radiative processes} come into play. Radiation may affect reconnection profoundly in several fundamental ways, none of which has been adequately explored so far. Perhaps the three most important radiation effects are: (a) prompt {\it radiative cooling} (optically thick or optically thin) of the reconnection layer; (b) {\it radiation pressure}; and (c) {\it radiative resistivity} (Compton drag). 

{\bf (a) Radiative cooling} due to various radiation mechanisms may be important in many astrophysical situations, e.g., (1) synchrotron and synchrotron-self-Compton in~GRB, AGN, and Blazar jets; (2) external inverse-Compton cooling of energetic electrons by powerful ambient soft radiation fields in coronae of black holes accreting at a large fraction of the Eddington limit, both in galactic X-ray sources and in AGNs;  (3) radiation diffusion out of an optically-thick pair-dominated reconnection layer in the context of magnetar flares and GRB central engines.

Radiative cooling has only recently started being included in reconnection models~\citep{Steinolfson_vanHoven-1984, Dorman_Kulsrud-1995, Jaroschek_Hoshino-2009, Giannios_etal-2009, Nalewajko_etal-2010, Uzdensky_McKinney-2010} and its effects on reconnection have only now started being studied in a systematic way, starting with a Sweet--Parker theory of non-relativisitc resistive reconnection in electron-ion plasma in the strong cooling regime~\citep{Uzdensky_McKinney-2010}. It was found that a strong prompt radiative cooling greatly affects the energy balance and hence the dynamics and thermodynamics of the reconnection layer. One of the most important effects of radiative cooling is that it limits the increase of the central temperature caused by ohmic heating and this, in turn, has a great effect on the plasma collisionality and Spitzer resistivity inside the layer: cooler plasmas are more collisional~\citep{Uzdensky_McKinney-2010}. In addition, in the absence of a strong guide field, radiative cooling leads to a strong plasma compression inside the layer to maintain the pressure balance~\citep{Dorman_Kulsrud-1995, Uzdensky_McKinney-2010}. 
Although the paper by~\cite{Uzdensky_McKinney-2010} concentrated mostly on optically thin cooling due to a number of radiative mechanisms (cyclotron, bremsstrahlung, inverse Compton), many of its ideas, concepts, and conclusions are valid more generally. However, I believe that any serious future efforts in this area will require approaching the reconnection problem as a radiative-transfer problem.

{\bf (b) Radiation pressure:} In some of the most spectacular high-energy astrophysics phenomena, the dissipated energy density and hence the plasma temperature are so high that radiation pressure (that scales as~$T^4$) starts to play an important role. Examples include reconnection in central parts of black-hole accretion disks; central engines and inner parts of GRB jets; and gamma-ray flares in magnetospheres of magnetars. In most extreme cases (magnetar flares and GRBs) radiation pressure (together with the relativistic pair pressure) is likely to completely dominate over the baryonic plasma pressure. How this affects the reconnection process is still not known.
 
{\bf (c) Radiative (Compton-drag) resistivity} due to collisions between electrons and photons
(in contrast to electron-ion collisions as in the classical Spitzer resistivity) is important in several high-energy astrophysics situations (accretion disk coronae of black holes, magnetar flares, and GRB jets).

To the best of my knowledge, the above fundamental aspects of magnetic reconnection in the presence of strong radiation have not yet been adequately explored so far, even though they are critical for a number of outstanding problems in modern high-energy astrophysics.  There is only a handful of papers on this subject \citep{Dorman_Kulsrud-1995, Uzdensky_MacFadyen-2006, Uzdensky-2008, Jaroschek_Hoshino-2009, Nalewajko_etal-2010}; (Uzdensky \& McKinney 2010) and clearly much more needs to be done. There is a clear astrophysical motivation for further effort in this fundamental research area, for investing resources in it to assure rapid progress. To some degree, this progress may be facilitated by the advent of new experimental and computational tools (e.g., radiation-MHD codes such as \cite{Farris_etal-2008}) developed recently in the emerging area of High-Energy-Density Physics (see below).

%--------------------------------------------------------------------------

{\bf (3) Pair Creation:} At the most extreme end of high-energy-density astrophysical reconnection lie environments such as magnetar magnetospheres and central engines of Gamma-Ray Bursts (GRBs) and Supernovae, as well as GRB inner jets. 
These systems are believed to possess ultra-strong magnetic fields ---  fields exceeding the critical quantum field of about $4 \times10^{13}$~G. The energy density of such a strong field is so high that, when converted to radiation, it yields equilibrium radiation temperatures greater than the electron rest-mass energy (0.5~MeV). 
In a situation like this, copious {\it electron-positron pair creation} inevitably occurs inside the reconnection layer~\citep{Uzdensky_MacFadyen-2006, Uzdensky-2008}, in addition to all the radiative processes mentioned above. 
The resulting great increase in plasma density makes the reconnection layer optically thick and highly collisional~\citep{Uzdensky_MacFadyen-2006}. To a certain degree, pair creation should play an important role also in somewhat lower-energy systems with mildly-subcritical fields ($10^{12}$~G), such as GRB jets at intermediate distances and normal neutron stars (e.g., in X-ray pulsars).

How radiation and pair production processes affect high-energy-density reconnection is still not clear.
This question will be the main topic of this paper, whereas special-relativistic effects will be left out to simplify the discussion.

%------------------------------------------------------

As the above discussion demonstrates, the additional physical complications that one encounters in astrophysical reconnection often have to do with the high {\it energy density} in these systems. Thus, a new, rich and exciting area of research --- magnetic reconnection in High-Energy-Density (HED) astrophysical environments --- lies wide open before us! 
One may hope that our efforts to understand the corresponding astrophysical phenomena will benefit from developing a better understanding of HED environments --- environments in which magnetic field, plasma, and radiation are all important and strongly coupled to each other. 
It is worth mentioning that HED Physics (HEDP) and Laboratory Astrophysics (HEDLA) represent a new and exciting branch of modern physics that has emerged in recent years~\citep[e.g.,][]{Drake-2006-book}; they study the properties of matter at very  high energy densities where radiation pressure is important.  HEDP is currently enjoying a period of rapid growth fueled in part by the availability of new powerful lasers and Z-pinches, such as NIF, Hercules, Vulcan, Omega,  Z-machine, and Magpie. There has also been a lot of progress on the computational side, with the development of advanced radiation-hydro and radiation-MHD codes. The advent of these new experimental and computational tools and capabilities provides new opportunities and calls for the exploration of fundamental plasma processes in the HED regime, including magnetic reconnection. Our understanding of radiative aspects of HED reconnection will especially benefit. 
So far, however, there has been only a handful of disjoint experimental (utilizing laser-produced plasmas with mega-gauss magnetic fields)~\citep{Nilson_etal-2006, Li_etal-2007} and theoretical~\citep{Uzdensky_McKinney-2010} investigations of various aspects of HED reconnection. I believe that a more coherent and organized effort in this important new area of fundamental research is needed.

%---------------------------------------------------------

%**************************************************************

\section{Reconnection in ultra-strong magnetic fields}
\label{sec-B_*}

%**************************************************************

In order to give the most dramatic illustration of the above physical effects that become important in HED astrophysical reconnection, this paper will focus on the most extreme case, where all of these effects (relativity, radiative effects, and pair creation) come into play in a very powerful way: reconnection of magnetar-strength fields. 
This problem, representing a new direction of reconnection research, has important astrophysical applications (magnetar magnetospheres, GRB central engines and jets).  In addition, it is intellectually attractive due to its rich and exotic physics. 

Before I start, I would like to remark that in traditional theoretical or computational reconnection studies one often likes to cast the problem in terms of dimensionless parameters and variables (e.g., velocity normalized to~$V_A$). When doing this, one  is making use of the fact that the equations one deals with can be rescaled, so that the actual absolute values of dimensional system parameters (magnetic field in Gauss or the system size in centimeters) do not matter. This approach is indeed justified for most of solar-system reconnecting systems. 
The problem we are going to consider here, however, is different. Here, the absolute scale of certain quantities, namely, the magnetic field, does in fact matter. There is an important natural scale for the magnetic field strength, a certain number in Gauss, and the physics of the problem depends on whether the reconnecting magnetic field~$B_0$ is low or high compared with this critical field. 

This important magnetic field scale is the critical quantum field, 
\beq
B_* \equiv  {{m^2 c^3}\over{e\hbar }}  \simeq 4.4\times 10^{13}\ {\rm G} \, .
\label{eq-B_*}
\eeq
This critical field has a clear physical interpretation that the electron cyclotron energy~$\hbar\Omega_e$ in this field is equal to the electron rest-mass energy. In this paper, I will be using the adjectives "magnetar-strength", "ultra-strong", and "super-critical" interchangeably, all meaning magnetic fields stronger than the critical field~$B_*$.

%--------------------------------------------------------------------

\subsection{Astrophysical Motivation}
\label{subsec-astro}

Although the physics of magnetic reconnection involving magnetic fields of~$4\times 10^{13}$~G and greater, including radiation and pair production, has not yet been seriously investigated and is very poorly understood, it is of great interest to astrophysics. Because magnetic fields of such high strength are relatively rare in the Universe, the scope of applicability of this problem is of course just a small (albeit, arguably, one of the most interesting and exotic!) subset of the very rich landscape of astrophysical reconnection painted in the Introduction. Specifically, there are two main astrophysical applications of ultra-strong field reconnection, both of which are hot topics in modern high-energy astrophysics: giant flares in magnetar (SGR) magnetospheres, and central engines and jets of GRBs and supernovae. I shall discuss these examples below. However, I also would like to remark that, in addition to these two main examples, radiation effects and pair production may become important even at somewhat lower reconnecting magnetic fields ($10^{12}$~G) encountered in magnetospheres of regular neutron stars in radio- and X-ray pulsars.

%--------------------------------------------------------------------

{\bf (1) Magnetar Giant Flares.}

Magnetars are isolated neutron stars with magnetic fields of order~$10^{15}$~G, much stronger than the magnetic fields typical for normal neutron stars ($\sim 10^{12}$~G) and higher than the critical quantum magnetic field~$B_*$.  
Observationally, magnetars are detected through their powerful X-ray and $\gamma$-ray emission~\citep{Woods_Thompson-2006}. Because they do not have binary companions, this emission cannot powered by accretion; instead, it is believed to be powered by the episodic dissipation of the magnetic field energy.
Most of the time, magnetars are found in a quiescent state, but sometimes they produce intense bursts of soft gamma-rays (so-called Soft Gamma Repeaters, or~SGRs). The most dramatic manifestation of magnetar activity is the SGR giant flares --- the most intense galactic events, releasing $10^{44}-10^{46}$~erg in just a fraction of a second~\citep[e.g.,][]{Mazets_etal-1999, Palmer_etal-2005}.
It is generally believed that the underlying mechanism for powering these flares is, again, the release of the free energy of a 
non-potential magnetic field in the neutron star's magnetosphere, twisted up by the star's crustal displacements~\citep{Thompson_Duncan-1995, Thompson_Duncan-2001, Thompson_etal-2002}. In particular, based on the analogy with solar flares, it has been suggested that the specific mechanism for magnetic energy release is reconnection in the magnetar's magnetosphere~\citep{Thompson_Duncan-1995, Thompson_Duncan-2001, Lyutikov-2003b, Lyutikov-2006a, Masada_etal-2010}. However, how exactly
reconnection should happen in this environement, e.g., how fast it should be, is not yet understood  and has not been systematically explored in the literature.
In view of this fact and in order to discriminate the magnetar reconnection hypothesis from competing alternative
models, it is extremely critical to understand the underlying physics of reconnection of magnetar-strength magnetic fields and to connect the theory with the key observational constraints: the short (milliseconds) flare rise time, and the duration (0.25~sec) and the temperature ($\sim 200$~keV) of the main $\gamma$-ray spike.

%----------------------------------------------------------------

{\bf (2) Magnetically-driven models of Gamma-Ray Bursts (GRBs).}

Long-duration gamma-ray bursts (GRBs) are some of the most energetic 
and spectacular explosions in the Universe, releasing on order
$10^{51}-10^{52}$~ergs of energy in the form of $\gamma$-radiation in just a few seconds or tens of seconds. 
The current paradigm, known as the collapsar model~\citep{Woosley-1993, Paczynski-1998,  
MacFadyen_Woosley-1999}, holds that these enigmatic explosions 
result from gravitational collapse of cores of massive stars.
In this scenario, a star's iron core collapses to form a  black hole, while the subsequently falling stellar 
material forms an accretion disk around it. The resulting  ``micro-quasar inside a star'' acts as the central 
engine driving the subsequent GRB explosion. 
Alternatively, the progenitor star's core may collapse into a rapidly rotating (millisecond) magnetar 
(instead of a black hole) that can also act as a central engine. 
In any case, the engine produces an ultra-relativistic jet that burrows through the outer stellar envelope, 
breaks through the star's surface, and eventually leads to the observed burst of gamma rays. 
In one of the leading versions of the collapsar model, the jet is launched and collimated magnetically 
\citep[e.g.,][]{Thompson-1994, Meszaros_Rees-1997, Lee_etal-2000, Vlahakis_Konigl-2001, 
Lyutikov_Blackman-2001, vanPutten_Ostriker-2001, Drenkhahn_Spruit-2002, 
Lyutikov_Blandford-2002, Lyutikov_Blandford-2003, vanPutten_Levinson-2003,
Proga_etal-2003, Lyutikov-2006b, Uzdensky_MacFadyen-2006, Uzdensky_MacFadyen-2007a, Uzdensky_MacFadyen-2007b, Burrows_etal-2007, Komissarov_Barkov-2007}, 
in analogy with jets in other astrophysical systems. 
Since most of the large-scale  poloidal magnetic flux is produced by MHD dynamo 
in the central engine (as opposed to the pre-existing field in the progenitor star), 
all the poloidal flux has to return to the central engine. Therefore, morphologically, 
the jet should be similar to a magnetic tower~\citep{Uzdensky_MacFadyen-2006}, introduced 
by~\cite{Lynden-Bell-1996} for the jets of active galactic nuclei (AGN). Recent MHD simulations 
of the collapsar and millisecond-magnetar scenarios seem to 
support this picture~\citep{Burrows_etal-2007, Komissarov_Barkov-2007}. 

In order to be successful, any magnetically-dominated GRB scenario
must have a magnetic field of order $10^{15}$~G at the base of the outflow 
--- the inner part (10~km) of the central engine's accretion disk or
the neutron star. 
This is in fact quite plausible for the dense and hot environment of a collapsing stellar core. 
Of key importance, however, is the question whether the resulting outflow will remain 
magnetically dominated as it propagates through the star. 
Indeed, a smooth coherent magnetic structure, e.g., a magnetic tower, is certainly an 
idealization, employed to obtain basic physical insight into the system's dynamics~\citep{Uzdensky_MacFadyen-2006}.
In reality, the magnetic field in the GRB jet may  have a complex substructure 
on smaller scales~\citep{Uzdensky_MacFadyen-2007a, Uzdensky_MacFadyen-2007b}.
Furthermore, the magnetic field in the jet (e.g., in a magnetic tower)
is constantly twisted up by the disk's differential rotation,
and thus may become kink-unstable. This may lead to a significant disruption, 
in particular, to the formation of thin current layers,
similarly to the internal-kink-driven sawtooth disruptions in tokamaks.
In addition, a large-scale current sheet naturally forms as the initially closed magnetic
field of the central engine opens up~\citep{Uzdensky-2002}. 

Current sheets formed by any of these processes are the possible sites for magnetic reconnection. 
If reconnection happens quickly enough, it may lead to the break up of a single coherent magnetic structure into a "train of plasmoids"~\citep{Uzdensky_MacFadyen-2007a}. The expected sizes of such plasmoids and their production rate are presently not known, but this should have important consequences for the observed rapid GRB variability. 

These considerations show that reconnection is an important issue for magnetically-driven GRB models. Whether a magnetic tower jet can survive as it propagates through the star, and whether it remains magnetically dominated at large distances, depends critically on the effectiveness of reconnection. In other words, reconnection onset determines where in the magnetically dominated GRB jet the magnetic energy is dissipated and this, in turn, has profound implications for observable radiative signatures including the prompt $\gamma$-ray emission~\citep{Lyutikov_Blackman-2001, Lyutikov_Blandford-2002, Lyutikov-2006b, Giannios_Spruit-2006, Giannios_Spruit-2007, McKinney_Uzdensky-2010}.

%**************************************************************

%--------------------------------------------------------------------------------------------

\subsection{Definitions and useful relationships}
\label{subsec-definitions}

As I mentioned above, although reconnection of magnetar-strength (of order $B_* \simeq 4\times 10^{13}$~G and greater) magnetic fields is often  invoked in astrophysics, its most basic underlying physics has not been systematically explored so far. Therefore, in order to approach this problem, one first needs to make preliminary estimates in order to figure out what the relevant physics is, what physical processes are important. This task will be the main objective of the present review. Only after this is accomplished, can one move on to the next stage and think about 
the most appropriate ways to describe the problem mathematically and explore simple toy models (e.g., along the lines of the Sweet-Parker reconnection model). Eventually, one hopes to gain enough insight from such models to be able to construct a physically realistic model that most likely will have to be attacked numerically.

Before I proceed, I would like to list, for reference, the following important expressions and definitions. 

Fine structure constant:
\beq
\alpha = {e^2\over{\hbar c}} \simeq {1\over{137}} \simeq 0.00723 \, .
\eeq

Classical electron radius:
\beq
r_e = {{e^2}\over{m_e c^2}} \simeq 2.818\times 10^{-13}\, {\rm cm} \, .
\eeq

Electron Compton wavelength (divided by~$2\pi$): 
\beq
l_C = r_e/\alpha = {\hbar\over{m_e c}} = 3.86\times 10^{-11}\, {\rm cm} \, .
\eeq

Thomson cross-section:
\beq
\sigma_T = {8\pi\over 3}\, r_e^2 \simeq 6.65\times 10^{-25}\, {\rm cm^2} \, .
\eeq

Stefan-Boltzmann constant:
\beq
\sigma_{\rm SB} = {\pi^2\over{60}}\, {k_B^4\over{\hbar^3 c^2}} \simeq 5.67 \times 10^{-5}\ {\rm erg\, cm^{-2}\,K^{-4}\, s^{-1}} \, .
\eeq

Radiation constant:
\beq
a = {4\, \sigma_{\rm SB}\over c} = {{\pi^2 \, k_B^4}\over{15}}\, (\hbar c)^{-3}  \simeq 7.566 \times 10^{-15}\ {\rm erg\, cm^{-3}\,K^{-4}} \, .
\eeq

Critical magnetic field, defined so that $\hbar\Omega_e \equiv \hbar eB_*/m_e c = m_e c^2$: 
\beq
B_* \equiv {{m_e^2 c^3}\over{e\hbar}} \simeq 4.41 \times 10^{13}\ {\rm G} \, .
\eeq

The magnetic pressure and the energy density of the critical magnetic field:
\beq
{B_*^2\over{8\pi}} = 
{1\over{8\pi}}\, (m_e c^2)^4\, \alpha^{-1}\, (\hbar c)^{-3}\, 
\simeq 7.8\times 10^{25}\, {\rm erg\ cm^{-3}} \, .
\eeq

%**************************************************************

\subsection{A remark about and large and very large numbers}
\label{subsec-remark}

In this paper we shall deal with approximations based on various
large dimensionless parameters. Some of these will be simply the
physical parameters of the problem, properly dimensionalized, such 
as the upstream reconnection field ($B_0$) in units of~$B_*$, the central temperature ($T_0$) in units of~$m_e c^2$, etc.
I shall sometimes formally consider $\theta_e \equiv k_B T_0/m_e c^2$ and~$b\equiv B_0/B_*$ to be large, 
but will keep in mind that in reality they are not tremendously large. 
In addition, we may encounter a few other dimensionless parameters that are just physical constants, e.g., the inverse fine structure constant~$\alpha^{-1}\simeq 137$. We shall also treat these formally as large numbers. 

Then, however, there is one truly large number in this problem: 
the ratio of the size of the system~$L$ to the electron Compton wavelength~$l_C$.
Indeed, keeping in mind astrophysical applications of our theory 
to neutron stars (magnetars, pulsars, core-collapse environments),
the typical system size is the radius of a neutron star (or, perhaps, the gravitational radius of a stellar-mass black 
hole in GRB case). Generally, it is of order~$10^6$~cm, and hence
$L/l_C \sim 10^{16}$. This number is much-much larger than all the other independent dimensionless numbers in the system. It therefore represents a different class of "largeness". 
As we shall see, several important large parameters in our problem can be expressed in terms of this ratio.

%------------------------------------------------------------------------------

%\subsection{Basic Physical ingredients of ultra-strong field reconnection}
%\label{subsec-ingredients}

%-----------------------------------------------------

\subsection{Relativistically hot plasma}
\label{subsec-temperature}

What are the physical challenges in understanding magnetic 
reconnection in the ultra-strong magnetic field regime?
Of key importance here is the fact that, when the corresponding magnetic energy is 
dissipated, the resulting internal plasma energy density and pressure are so 
high that they become dominated by radiation (and, for $B_0\gg B_*$, by 
relativistic pairs as well). In particular, an important basic estimate for the temperature~$T_0$ at the center of the reconnection layer in the case without a guide magnetic field can be readily obtained from the cross-layer pressure balance. Assuming for simplicity that the region outside (upstream) the layer is magnetically dominated, so that the upstream plasma and radiation pressure can be neglected, the pressure balance condition can be written as
\beq
 P_{\rm rad, 0} + P_{\rm plasma, 0} = {{B_0^2}\over{8\pi}} \, ,
\label{eq-pressure-balance}
\eeq
where $B_0$ is the upstream magnetic field, $P_{\rm rad, 0}$ is the radiation pressure and $P_{\rm plasma, 0}$ is the plasma pressure, including pairs, at the center of the reconnection layer. 

In what follows it will be convenient to represent the temperature by a dimensionless quantity --- a temperature normalized by the electron rest-mass energy:
\beq
\theta_e \equiv {k_B T_0\over{m_e c^2}}  \, .
\label{eq-theta-def}
\eeq

Then, the corresponding radiation pressure is
\beq
P_{\rm rad} = {1\over 3} \, U_{\rm rad} = {a\over 3}\,  T^4 = 
{{\pi^2}\over{45}}\, (m_e c^2)^4\, (\hbar c)^{-3}\, \theta_e^4 =
{{\pi^2}\over{45}}\, {{m_e c^2}\over{l_C^3}} \, \theta_e^4 \, .
\eeq

For the extreme conditions we are interested in here, we expect the pressure of baryons and associated electrons inside the layer to be small compared with the radiation and relativistic pair pressure. The pair pressure $P_{\rm pairs}$ in thermal equilibrium is lower than the radiation pressure for moderate temperatures, $T\leq m_e c^2$, but becomes comparable to the radiation pressure at higher temperatures, $P_{\rm pairs} \simeq (7/4) \, P_{\rm rad}$ \citep[e.g.,][]{Choudhuri-2010}. Thus, we can approximate the total gas+radiation pressure inside the layer by the $P_{\rm rad}+P_{\rm pairs} \simeq (11/4)\, P_{\rm rad}$, and thus get fiducial estimate for the central layer temperature~$T_0$  based on pressure balance between the magnetic pressure outside and the radiation and pair pressure inside:
\beq
\theta_{e0} \equiv {k_B T_0\over{m_e c^2}} = \biggl({4\over 11}\, {45\over{8\pi^3\alpha}}\biggr)^{1/4}\, b^{1/2}
\simeq 0.507\, \alpha^{-1/4}\, b^{1/2} \simeq 1.73\, b^{1/2} \, .
\label{eq-theta_0}
\eeq
where 
\beq
b \equiv B/B_* \,  .
\eeq
In any case, we see that we are necessarily dealing with relativistically hot plasmas!
Basically, we see that the central plasma is relativistically hot for $b> 1$ and not relativistic for $b< 1$. 
In particular, for a typical magnetar field of $B_0=10^{15}\ {\rm G} \simeq 23\, B_*$, we get $k_B T\simeq
4\, {\rm MeV} \simeq 8\, m_e c^2$.

Interestingly, because radiation and pair pressures, which are the dominant pressure components in this regime, 
depend only on temperature but not on the baryonic density, the latter drops out of the pressure balance. Hence, the pressure balance uniquely determines the temperature in terms of the upstream magnetic field, essentially independently of the upstream plasma density!

%----------------------------------------------------------------------------------------

\subsection{Pair Creation}
\label{subsec-pairs}

What are the implications of this very high temperature inside the layer?
The main implication is that the temperature inside the layer is so high 
that one expects copious electron-positron pair production. In the case 
of magnetar flares, the pair density inside the layer is likely to be rather insensitive to, and 
much greater than, the ambient plasma density. In fact, it is determined essentially 
uniquely by the strength $B_0$ of the reconnecting magnetic field, no matter how 
low the upstream plasma density is~\citep{Uzdensky_MacFadyen-2006}.
In particular, in the ultra-relativistic limit  $\theta_{e0}\gg 1$ (corresponding to $b\gg 1$), 
the equilibrium pair density  is
\beq
n_{\rm pairs} \simeq {{3\zeta(3)}\over{2\pi^2}}\, \bigg({T\over{\hbar c}}\biggr)^3 
\simeq 0.1827\, \bigg({T\over{\hbar c}}\biggr)^3 \simeq
0.1827\ l_C^{-3}\, \theta_e^3  \simeq 3.2\times 10^{30}\ {\rm cm}^{-3}\,  \theta_e^3 \, ,
\label{eq-n_pairs}
\eeq
where $\zeta(x)$ is Riemann's zeta-function, $\zeta(3)= 1.202...$.
Thus, we see that the characteristic scale for the pair density in this regime is one particle per~$l_C^3$, and for $B_0=10^{15}$~G the density becomes as high as $3\times 10^{33}\ {\rm cm}^{-3}$ --- again, much higher than, and hence essentially independent from, the ambient plasma density in the magnetar magnetosphere. 
This justifies out our assumption that the pressure inside the layer should be dominated by pairs and radiation, whereas the baryonic contribution should be negligible. 

As a side remark, it is interesting to note that, as was shown by \cite{Medvedev-1999}, the plasma density $\omega_p$ in an ultra-relativistic pair plasma in thermodynamic equilibrium with photons, is just proportional to the temperature:
$\hbar \omega_p/k_B T = [14 \zeta(3) \alpha / 3 \pi^2]^{1/2} \simeq 0.0644$. Thus, for example, for $k_B T_0= 4\, {\rm MeV}$, corresponding to $B_0\simeq 10^{15}\, {\rm G}$, we get $\hbar \omega_p \simeq 260\, {\rm keV}$.

%----------------------------------------------------------------------------------------

Why are pairs important? 
They are important because they are expected to affect in the physics of reconnection in two profound ways. 
Firstly, the pairs make the reconnection layer optically-thick even in the direction {\it across} the layer,
hence efficiently trapping the radiation and ensuring local thermodynamic equilibrium between the plasma and radiation in the layer. Secondly, the extremely high pair density makes the plasma inside the layer highly collisional~\citep{Uzdensky_MacFadyen-2006}, in contrast to reconnection in tokamaks, in the Earth magnetosphere, and, to some degree, in solar flares.
We will  consider these two consequences of pair production in relativistically hot reconnection layer 
in the next two subsections.

%--------------------------------------------------------------------------

\subsection{Optical depth and the radiative transfer problem}
\label{subsec-opt-depth}

First I would like to note that, in the case of reconnection with supercritical fields and highly-relativistic central layer temperatures, we expect that a high pair density should not be limited to the current layer proper, but may, in fact, spread over a much wider region, forming a thick "pair coat" around the current layer. 
 
Indeed, as the magnetic energy is dissipated and the reconnection layer heats up, it starts to emit large numbers of $\gamma$-ray photons, including those with energies  $\epsilon> m_e c^2$. These photons create electron-positron pairs some distance upstream of the current layer. Some of the newly created pairs quickly produce new $\gamma$-ray photons via pair annihilation and other mechanisms and some of these new photons in turn travel some distance away from the layer and then produce new pairs, etc. Thus, the thin  resistive current layer, in which the ohmic dissipation of magnetic energy takes place, quickly becomes {\it dressed} in a growing {\it coat of electron-positron pairs}. 

Pair creation is affected by quantum-electrodynamical (QED) effects associated with the ultra-strong magnetic field through which the photons have to propagate. On the one hand, the strong magnetic field suppresses somewhat the usual two-photon pair creation process, but, on the other hand, it enables a new process of one-photon pair creation~\citep{Harding_Lai-2006}. In any case, however, our basic conclusion about the existence of a substantial pair coat surrounding the current layer is likely to remain valid.

The column density of the pair coat is likely to be large enough to efficiently trap the photons, even in the direction {\it across} the layer; that is, the pair coat is expected to be optically thick to Compton/Thomson scattering.  The high optical depth ensures a good local thermal coupling between the photons and the plasma. This expectation alone represents a stark difference with the conventional reconnection where the reconnection region is always considered to be completely optically thin.

In addition to a large optical depth to Compton scattering on pairs, the ultra-strong magnetic field itself provides a substantial obstacle to the propagation of photons due to QED processes of 1-photon pair creation ($\gamma\rightarrow e^+ e^-$) and photon-splitting ($\gamma \rightarrow \gamma\gamma$). These processes are especially important for higher-energy photons and effectively preclude most of MeV photons from leaving the inner magnetosphere. 

The pair coat may also be geometrically significantly thicker than the current layer proper, and thus may appear to an outside observer as a moderately flattened "fireball" with some photospheric temperature~$T_{\rm ph}$. 

In general, because the equilibrium pair density, resistivity, and radiation pressure are all strong functions of temperature, the issues of thermodynamics and thermal transport are extremely important in this problem. In particular, one needs to  pay careful attention to the balance between ohmic heating, advection, and radiative cooling. 

Because of the large expected optical depth, photons have to diffuse gradually across the dressed layer. 
Nevertheless, under certain conditions, the corresponding photon diffusion time may be shorter than the global advection time along layer. 
Specifically, for a dressed layer of length~$L$ and thickness~$H < L$ and scattering optical depth~$\tau$, the photon diffusion time can be estimated as $t_{\rm diff} \sim \tau H/c$, whereas the global advection time is about $t_{\rm adv} \sim L/c$. Thus, photons can escape the layer before being advected out by the plasma flow if $\tau < L/H$, or, equivalently, if $\tau^2 \sim L/\lambda_{\rm ph, mfp}$. That is, unless the optical depth is too large, a significant part of the dissipated magnetic energy can be promptly radiated away across the layer, instead of being advected out along the layer. 
In this {\it strong cooling} regime, radiative cooling competes with (or even dominates over) advection as the main mechanism of removing the dissipated energy. Intense radiative cooling governs the thermodynamics of the reconnection process and drastically affects both the dynamics of reconnection layer and its appearance to an external observer. 

Importantly, in the strong cooling regime our magnetic reconnection problem becomes, fundamentally, 
a {\it radiative transfer problem}! 
This, again, is in sharp contrast with conventional low-energy-density reconnection (e.g., in solar flares, Earth magnetosphere, and tokamaks), where radiative energy losses are insignificant on the global advection time scale.
Furthermore, in the strong cooling case, the two directions (across and along the layer) effectively decouple, 
which makes  the analysis much easier. In particular, the problem of determining the perpendicular layer structure becomes one-dimensional (1D) and involves two coupled equations: the pressure balance and the radiative transfer equation reflecting the balance between ohmic heating and radiative cooling (in many respects, this is similar to the problem of the vertical structure of accretion disks). No such simplification is possible in the traditional reconnection case where the dissipated energy is removed by the flow of plasma along the layer and the problem remains essentially two-dimensional.

Despite being one-dimensional, the radiative transfer problem in the strong cooling regime is still highly nontrivial, in part because it needs to be coupled with an equation for the pair density. Deep below the photosphere, the plasma-radiation thermal coupling is good and the pair density is determined by the local thermodynamic equilibrium [e.g., by an analog of the Saha equation, see eq.~(\ref{eq-n_pairs})]; it essentially becomes a function of local temperature only. However, near and certainly outside the photosphere, local thermodynamic equilibrium breaks down and the pair density is instead determined by the balance between various creation and annihilation reactions, which are all affected by the ultra-strong magnetic field through a number of nontrivial QED effects. 

This problem (even below the photosphere) is yet to be solved; we don't even have a good theoretical handle on the most basic  parameters of the "dressed" reconnection layer. 
Moreover, in its full generality (i.e., for arbitrary optical depth), the problem of reconnection with strong radiative cooling is not yet solved even for a sub-critical magnetic field, $B\ll B_*$, when pair creation is not important. A first step towards solving it  was made recently by~\cite{Uzdensky_McKinney-2010} who investigated the effects of strong radiative cooling on resistive-MHD reconnection with a main focus on the optically thin regime. For simplicity, that paper considered a non-relativistic compressible resistive (with Spitzer resistivity) reconnection layer and ignored radiation pressure and Compton-drag resistivity. It formulated the conditions for the strong-cooling regime and developed a self-consistent Sweet--Parker-like model with strong radiative cooling. In this regime, the temperature at the center of the layer is determined by the balance between ohmic heating and radiative cooling and is substantially lower than in the case without cooling. Correspondingly, because of the strong temperature dependence of the Spitzer resistivity, the Lundquist number $S\propto 1/\eta$ is lower and the reconnection rate is higher. Furthermore, in the case of antiparallel reconnection (i.e., without a guide field), intense cooling leads to a strong compression of the plasma inside the layer and this also enhances the reconnection rate. 

Although the main focus of~\cite{Uzdensky_McKinney-2010} was on the {\it optically thin} cooling regime for several specific astrophysically important radiation mechanisms, many of the results and insights obtained in that paper are quite general and should apply to the optically thick and intermediate cases as well. 
However, these results are not directly applicable to the high-energy-density, radiation- and pair-dominated environment we are considering here, for a number of reasons. One of them is that the relativistic pair and radiation pressure, which dominates the pressure balance in our problem, is insensitive to the baryonic density and hence it is not clear whether a strong radiative cooling needs to lead to strong baryonic compression. 
And, of course, the problem of reconnection of magnetar-strength fields is further complicated by an entire host of pure-QED effects that govern the propagation of high-energy photons across the strong magnetic field, as well as pair creation and annihilation processes. Among the most important such effects are magnetic-field suppression of Compton scattering and of two-photon pair creation, one-photon pair creation process, and photon splitting [see \cite{Harding_Lai-2006} for a review].

%---------------------------------------------------------------------------------------------

Interestingly, however, even though the full 1D structure of the dressed reconnection layer is not yet known, one can derive an important relationship between two key parameters of the reconnection layer in the strong optically-thick radiative cooling regime without guide field and assuming a steady state. These parameters are the optical depth across the dressed layer, $\tau$, and the reconnection rate~$E$. The resulting relationship appears to be rather robust and independent of the particular solution details. 

In the strong radiative cooling case, where a significant fraction of the dissipated magnetic energy is radiated away across the layer, the photospheric temperature~$T_{\rm ph}$ of the pair coat can be determined from the energy balance:  
\beq
\sigma_{\rm SB}\, T_{\rm ph}^4 \sim S_{\rm Poynt} \equiv {c\over{4\pi}}\, E B_0 = {v_{\rm rec}\over{4\pi}}\,  B_0^2 \, . 
\eeq
where $S_{\rm Poynt}$ is the Poynting flux coming into the layer. 
On the other hand, assuming that there is no guide field, the central temperature $T_0$ of the layer is determined by the pressure balance between the magnetic pressure outside the layer and the radiation and pair pressure inside [see~eq.~(\ref{eq-pressure-balance})]:
\beq
P_{\rm rad,0} + P_{\rm pairs,0} = {11\over 4} \, {a\over 3}\, T_0^4 = {B_0^2\over{8\pi}}\, .
\eeq

Thus, we immediately get a simple relationship between the optical depth across the dressed layer, $\tau \simeq T_0^4/T_{\rm ph}^4$ and the dimensionless reconnection rate:
\beq
\tau  \simeq  {B_0\over{E}} \sim {c\over{v_{\rm rec}}} \, ,
\label{eq-tau-E}
\eeq
where we ignored factors of order unity.
That is, the thicker the layer, the slower is the reconnection process.

This result is interesting because one can actually try to test it against observations.
Indeed, the typical reconnection inflow velocity is related to the duration $t_{\rm rec}$ of the reconnection event
\beq
t_{\rm rec} \sim L/ v_{\rm rec} \, , 
\label{eq-t_rec-v_rec}
\eeq
where $L$ is the size of the reconnecting region, typically of order the size of the neutron star, $L\sim R_{\rm NS} \sim 10^6\, {\rm cm}$. The reconnection time $t_{\rm rec}$ presumably corresponds to the duration of the main prompt phase of giant magnetar flares: $t_{\rm rec} \sim t_{\rm flare} \sim 0.3\ {\rm s}$ \citep{Mazets_etal-1999}.  We thus expect $v_{\rm rec}/c \sim 10^{-4}$. 

On the other hand, the observed photospheric temperature during the prompt flare stage is of order 250~keV~\citep{Mazets_etal-1999}, 
$\theta_{\rm ph} = 0.5$, and so, taking the reconnecting field to be $B_0 = 4\times 10^{14}\, {\rm G} \simeq 10\, B_*$, 
the corresponding optical depth is
\beq
\tau = {T_0^4\over{T_{\rm ph}^4}}  \simeq 250\, b^2  \sim 2.5\times 10^4 \sim c/v_{\rm rec} \, ,  
\eeq
consistent with our expectation~(\ref{eq-tau-E}) and the above estimate $v_{\rm rec}/c \sim 10^{-4}$ based on giant flare observations. 

We note, however, that there is nothing really profound in the above result.  A similar result would be obtained for any 
optically-thick fireball, and so the apparent agreement between the expected optical depth and the reconnection rate cannot be viewed as a proof that reconnection is indeed the underlying mechanism for giant magnetar flares. Moreover, the optical depth only enters the above estimates indirectly, and what we are really comparing there are the observed photospheric temperature and the reconnection time scale. Thus, the above argument should be regarded just as a consistency check. 

Furthermore, in reality, it is actually very difficult to imagine a photosphere as hot as 250~keV. The reason for this is that the large number of pair-producing photons being emitted from such a photosphere would create more pairs, thereby growing the dense pair coat to even larger optical depths until a much lower (of order 20~keV as discussed later) photospheric temperature is reached. Thus, I concur with the traditional point of view that the apparent very high photospheric temperature of order 250~keV can only be explained by relativistic beaming of a much colder (20~keV) photosphere moving towards the observer with a Lorentz factor of order~10.

%--------------------------------------------------------------------

In order to get a deeper and more accurate understanding of how and where the photosphere forms, one needs to analyze the propagation of photons of different energies and polarizations across the outer layers of the pair coat and across an ultra-strong magnetic field. Photon propagation is greatly complicated by the presence of the ultra-strong field, but, overall, the picture can be summarized briefly as follows [see, e.g., \cite{Harding_Lai-2006} for review]. 

First, one needs to distinguish between the two photon polarizations: the ordinary mode (the O-mode), whose electric field vector has a component parallel to the background magnetic field, and the extraordinary mode (the X-mode), whose electric field is perpendicular to the background magnetic field. 
The propagation of X-mode photons across the field is strongly suppressed by the QED process of photon splitting. 
The attenuation coefficient for this process becomes insensitive to $B$ for $B> B_*$, but is a strong function of the photon energy ($\sim \epsilon^5$). Basically, only photons with $\epsilon < 0.1\, m_e c^2 \simeq 50\, {\rm keV}$ are able to escape the strong-field region of size $L\sim R_{\rm NS} \sim 10^6\, {\rm cm}$. In addition, X-mode photons with energies higher than the threshold for pair creation with one particle created at zeroth and the other at the first excited Landau level [this threshold is $\epsilon = (1+\sqrt{1+2b})\,m_e c^2 \simeq 3 \, {\rm MeV}$ for $b=10$] are subject to strong attenuation due to 1-photon pair production. For these pair-producing photons this process is even more important  than photon splitting. In addition, X-photons are subject to Compton scattering, which, however, is strongly suppressed by the magnetic field: $\sigma_{\rm es}^{(X)}/\sigma_T \simeq (\epsilon/b\, m_e c^2)^2$ for a photon with energy $\epsilon \ll b\, m_e c^2$ propagating perpendicular to the magnetic field~\citep{Herold-1979}.

In contrast, O-mode photons are not subject to photon splitting. Sub-MeV O-photons thus propagate essentially freely across the magnetic field; the only source of opacity for them is Compton scattering on pairs, which is not strongly suppressed by the magnetic field. On the other hand,  pair-producing (MeV) photons are subject to heavy absorption due to 1-photon pair production. This process and its inverse essentially ensure that these MeV photons are in equilibrium with pairs. 

Since, as we see, the propagation of MeV photons in a super-critical magnetic field is complicated, I shall concentrate here on the photosphere for sub-MeV photons that do not participate in pair production. After all, it is these photons that account for most of the observed emission in giant magnetar flares. 

The photosphere for sub-MeV O-photons is formed at a surface from which these photons are able to fly away freely, without being scattered by pairs. This means that the  photon mean free path becomes comparable with either the characteristic length scale of its own gradient, or the system size~$L$ (because at distances larger than $L$ the slab geometry is no longer applicable and we get geometric dilution effects). For simplicity, let us focus on the latter condition. 
Then, the photospheric temperature $\theta_{\rm ph}$ for sub-MeV photons is determined by a simple algebraic equation: 
\beq
 \lambda_{\rm ph,<MeV}^{\rm (O)} (\theta_{\rm ph})  = L \, .
\label{eq-photosphere-condition}
\eeq
 
Now, how does one estimate the mean free path of sub-MeV photons in the outer parts of an optically thick dressed layer?
Because of the very tight collisional coupling between pairs and $\geq$~MeV photons, facilitated by the ultra-strong magnetic field,  there is an approximate local thermodynamic equilibrium between pairs and such photons. Thus, as long as we are below or close to the photosphere for MeV photons, the local pair density can probably be determined from the Saha equilibirum formula.  
We expect the temperature at the photosphere (for both sub-MeV and MeV photons )to be somewhat sub-relativistic, so we can use the low-temperature limit of that formula with the exponential suppression of the pair density~\citep[e.g.,][]{Padmanabhan-2000}.
%[e.g., Padmanabhan, ``Theoretical Astrophysics'', vol.~I, p. 239-240, eq.~(5.239)]. 
However, we still expect the pair density to be much higher than the ambient baryonic density, which is therefore neglected. We then estimate: 
\beq
n_{\rm pairs}(\theta_e) \sim {1\over{\sqrt{2\pi^3}}}\, l_C^{-3}\, \theta_e^{3/2}\, e^{-1/\theta_e}\, .
\eeq

%On the other hand the density of MeV photons is, according to the Planck distribution, 
%\beq
%n_{\gamma,\rm MeV}(E) dE \simeq {2 E^2\over{\pi \hbar^2 c^3}}\, e^{-E/{kT}} dE/(2\pi\hbar)
%\eeq
%Then, the number density of photons capable of pair-producing is roughly, 
%\beq
%n_{\gamma,\rm MeV} \sim {2 (m_e c^2)^2\over{\pi \hbar^2 c^3}}\, e^{-1/\theta_e}\, kT/ (2\pi\hbar)
%\sim {{(m_e c^2)^3}\over{ \hbar^3 c^3}} \theta_e  \, e^{-1/\theta_e} = \lambda_C^{-3}\theta_e  \, e^{-1/\theta_e}
%\eeq
%
%Thus, the number of MeV photons in this regime is higher than the number of pairs by a factor
%\beq
%{{n_\pm}\over{n_{\gamma,\rm MeV}}} \sim \theta_e^{1/2}
%\eeq
%

In the presence of an ultra-strong magnetic field, the cross-section for the Compton scattering is no longer equal to Thomson cross section (with Klein-Nishina corrections) but becomes a nontrivial function of the photon energy and orientation; basically, a strong magnetic field suppresses Compton scattering somewhat. For low-energy O-photons propagating across the magnetic field, this suppression is rather modest, however~\citep{Herold-1979, Harding_Lai-2006}.  Thus, since here we are interested only  in the overall qualitative picture and in crude estimates, we can take it to be~$\sigma_T$. 
Then, the resulting Compton/Thomson mean free path for sub-MeV O-mode photons can be estimated roughly as
\beq
\lambda_{\rm ph, <MeV}^{\rm (O)} (\theta_e) \sim {1\over{\sigma_T \, n_{\rm pairs}}} \sim 
 l_C\, \alpha^{-2}\, \theta_e^{-3/2}  \, e^{1/\theta_e} \, .
\eeq

Substituting this into~(\ref{eq-photosphere-condition}), we see that  the photospheric temperature is determined from the algebraic equation
\beq
 \theta_{\rm ph}^{-3/2}\, e^{1/\theta_{\rm ph}} = \alpha^2\,  L/l_C \, . 
\eeq
Anticipating $\theta_{\rm ph}\ll 1$, its approximate solution is
 \beq
 \theta_{\rm ph}^{-1}  = \ln{\alpha^2 L/l_C} - {3\over 2}\,\ln\theta_{\rm ph}^{-1} \simeq  \ln{L/l_C} + 2\ln\alpha - {3\over 2}\,\ln\ln{L/l_C} \sim 25 \, .
 \eeq
where we took $L = 10^6 \ {\rm cm} \sim 2\times 10^{16} \, l_C$. As we see the resulting temperature depends on~$L$ and other parameters only logarithmically,  so the exact value of~$L$ is not important. Thus, we see that the photosphere for sub-MeV O-photons 
forms at a temperature of about 20~keV, and is rather insensitive to the exact parameters. We note that this temperature does not agree with the temperature ($T\sim 200\ {\rm keV}$ \citep{Mazets_etal-1999} observed during the main prompt (spike) part of an SGR flare lasting a fraction of a second. Thus, the very high observed  temperature of that spike can probably be explained only by invoking special-relativistic effects, namely an ultra-relativistic (Lorentz factor of~$\sim 10$) outflow beamed towards the observer.
On the other hand, the above photospheric temperature of $T_{\rm ph} \sim 20\, {\rm keV}$ is consistent with the observed thermal radiation emitted during the second (tail) phase of an SGR flare that follows the main spike and lasts for hundreds of seconds. This phase is conventionally explained by thermal radiative cooling of a highly optically thick relativistically-hot fireball confined magnetically next to the neutron star. Our optically thick pair coat around a reconnection layer is in some respect just a particular (flattened) version of such a fireball; thus, the model proposed in this paper can be viewed as providing a particular mechanism for the formation of such a fireball. 

Incidentally, the total dressed-layer optical depth corresponding to a photospheric temperature of 20~keV, 
and the central layer temperature of 2.6~MeV [which corresponds to the reconnecting magnetic field of $10\,B_*$, see
equation~(\ref{eq-theta_0})], can be estimated as 
\beq
\tau = T_0^4/T_{\rm ph}^4 \simeq 10\,b^2\, [\ln (\alpha^2\,L/\lambda_C)]^4 \sim  6\times 10^6\, b^2 \sim 10^9 \, .
\eeq 
We see that the expected reconnection time scale based on the strong-cooling formula~(\ref{eq-tau-E}) and equation~(\ref{eq-t_rec-v_rec}) would be about $t_{\rm rec} \sim \tau L/c \sim 3\times 10^4 \, {\rm sec}$, inconsistent with the observed flare duration. The reason for this is that the above optical depth is so large that the photon diffusion time across the layer is now {\it longer} than the global advection time along the layer, and so either the layer is no longer in the strong-cooling regime or the steady state assumption breaks down. In any case, the formula~(\ref{eq-tau-E}) is no longer applicable.  This means that here we are dealing with a more conventional situation where most of the dissipated magnetic energy stays with the plasma during the reconnection process. We then arrive at the traditional picture analogous to a solar post-CME flare~\citep{Thompson_Duncan-1995, Thompson_Duncan-2001, Lyutikov-2006a, Masada_etal-2010}. In this picture, one relativistically hot fireball is trapped in the post-flare reconnected magnetic loops attached to the neutron star and is responsible for the long tail thermal (with $T\sim 20$~keV) emission lasting for hundreds of seconds and modulated by the star's rotation, while another hot plasma blob is ejected out relativistically and is responsible for the main $\gamma$-ray spike.

%*************************************************************************

\subsection{Reconnection layer collisionality}
\label{subsec-collisionality}

The second major effect that pairs have on reconnection is their effect on the plasma {\it collisionality}:
the expected number density of pairs and photons inside the reconnection layer is so high that the layer becomes highly 
collisional~\citep{Uzdensky_MacFadyen-2006, Uzdensky-2008}. Here, the term "collisional" is used in the same sense as it is used in the traditional reconnection studies. Basically, it means that the corresponding Sweet--Parker~\citep{Sweet-1958, Parker-1957}  thickness $\delta_{\rm SP}= \sqrt{L\eta/V_A}$ is larger than the microphysical length scales relevant to collisionless reconnection, namely, the ion-collisionless skin depth~$d_i$ in electron-ion plasmas \citep[e.g.,][]{Ma_Bhattacharjee-1996, Biskamp_etal-1997, Rogers_etal-2001, Cassak_etal-2005} and the electron collisionless skin depth~$d_e$ in pair plasmas \citep[e.g.,][]{Bessho_Bhattacharjee-2005, Bessho_Bhattacharjee-2007} [replaced by the ion-sound \citep[e.g.,][]{Kleva_etal-1995, Rogers_etal-2001, Cassak_etal-2007} and electron Larmor radii, respectively, for the case with a strong guide magnetic field]. As one can easily show, the condition $\delta_{\rm SP} > d_i$ in the case without a strong guide field can be expressed in a form that brings out the role of collisionality explicitly: $L > \lambda_{e,\rm mfp} \, (m_i/m_e)^{1/2}$, where $\lambda_{e,\rm mfp}$ is the electron collisional mean free path inside the reconnection layer~\citep{Yamada_etal-2006}. Furthermore, for Coulomb collisions, this condition can be recast in an even more convenient form that involves only macroscopic system parameters~\citep{Uzdensky-2007}.

The reason why the layer collisionality is usually considered to be important is that it is generally believed to control the regime of reconnection and hence the reconnection rate, at least in traditional non-relativistic electron-ion plasmas. Namely, in the collisionless case resistive-MHD approximation breaks down inside the layer and one has to invoke more realistic plasma physics, especially in the generalized Ohm law. Then, as numerous numerical and laboratory studies have shown (\citet[e.g.,][]{Birn_etal-2001, Yamada_etal-2010}), reconnection proceeds in the so-called fast collisionless regime, 
facilitated by two-fluid (e.g., Hall) effects and/or anomalous resistivity and exhibiting a Petschek-like structure and a very fast reconnection rate, $cE\sim 0.1\, B_0 V_A$. In contrast, in the collisional ($\delta_{\rm SP} > d_i$) case, resistive-MHD description with classical resistivity is applicable; then, the two-fluid and anomalous-resistivity effects are not important, the \cite{Petschek-1964} fast reconnection mechanism fails, and reconnection proceeds much slower. 
The naive expectation would be that collisional resistive-MHD reconnection proceeds as slow as at the classical Sweet--Parker rate, $cE\sim B_0 \, (V_A \eta/L)^{1/2}$ (Sweet, Parker). However, recent studies have challenged this traditional point of view and blurred the distinction between the slow collisional and the fast collisionless reconnection. It now appears that even collisional reconnection may be fast! The reason for this is that 
thin and long resistive current layers are subject to a secondary tearing instability~\citep{Loureiro_etal-2007} that quickly breaks them up into chains of rapidly growing plasmoids~\citep{Samtaney_etal-2009}. As a result, a stationary and laminar reconnection layer is replaced with a highly dynamic hierarchical plasmoid chain~\citep{Shibata_Tanuma-2001, Uzdensky_etal-2010, Huang_Bhattacharjee-2010} with an effective reconnection rate of order $cE\sim 10^{-2}\, B_0 V_A$ even in resistive MHD~\citep{Bhattacharjee_etal-2009, Cassak_etal-2009, Huang_Bhattacharjee-2010, Uzdensky_etal-2010}. Furthermore, the plasmoid hierarchy leads to the development of very small current sheets, as thin as~$\delta_c \sim 100\, \eta/V_A$~\citep{Daughton_etal-2009, Huang_Bhattacharjee-2010, Uzdensky_etal-2010}.  If this thickness turns out to be smaller than~$d_i$, then these layers may themselves become collisionless~\citep{Daughton_etal-2009,  Shepherd_Cassak-2010, Uzdensky_etal-2010}. This leads to a certain blurring of the boundary between the collisional and collisionless reconnection regimes.  In particular, one expects the effective reconnection rate in this case to be nearly as high as  the canonical collisionless rate of~$0.1\, B_0 V_A$. 

I would like to remark that, until just a couple years ago, most of the traditional reconnection research in the past two decades has focused on collisionless reconnection. The main reason for this was that the main applications driving reconnection research --- reconnection in the Earth magnetosphere, sawtooth crashes in tokamaks, and, to a certain degree, solar flares --- do indeed involve sufficiently tenuous, collisionless environments. In addition, as we discussed above, resistive-MHD reconnection was expected to result only in very slow, Sweet--Parker reconnection, and hence to be irrelevant for any observed phenomena --- although we now know this not to be the case.  

In any case, while the jury on the rate of collisional reconnection may still be out, it is probably still true that the plasma collisionality inside the reconnection layer controls the regime of the reconnection and the overall reconnection-region structure and may also control the reconnection rate. Therefore, it is probably wise to pay a serious attention to this potentially important issue.  Of course, there is no guarantee that our intuition from non-relativistic, non-radiative plasmas can be of any use in the case of magnetar-strength field reconnection, but it is still worth asking whether the reconnection of magnetar-strength magnetic fields is collisional or collisionless.  

I claim that in the case of reconnection of super-critical (and even mildly subcritical) magnetic fields the electron-positron pair density should be so high that the reconnection layer should unavoidably be highly collisional (from the point of view of reconnection, as discussed above). To illustrate this, I present, with slight modifications, a specific example considered by~\cite{Uzdensky_MacFadyen-2006} in the GRB central engine context. 

For a reconnecting magnetic field of only about one half of~$B_*$, the corresponding plasma temperature at the center of the reconnection layer should be of order $k_B T_0\sim  300~{\rm keV}$. The corresponding equilibrium pair density [using the non-relativistic expression $n_e=2(m_e k_B T_0/2\pi\hbar^2)^{3/2}\,\exp(-m_e c^2/k_B T_0)$ for a rough estimate] is on the order of~$n_e\sim 2\times 10^{29}~{\rm cm}^{-3}$.
The classical collisional Spitzer resistivity due to electron-positron collisions at the above temperature is~$\eta \simeq 0.27 \, c r_e \theta_e^{-3/2} \ln\Lambda \sim 0.1~{\rm cm}^2/{\rm sec}$, and the Compton-drag resistivity should give a similar contribution at these temperatures. Then, the main dimensionless quantity that characterizes a resistive-MHD reconnection layer --- the Lundquist number $S\equiv V_A L/\eta$ --- is indeed very large, roughly, $S\sim (L/r_e) \sim 3\times 10^{18}$. Here, we took the global 
length-scale of the reconnecting system to be $L=10^6~{\rm cm}$ and used~$V_A\sim c$, $\theta_e \sim 1$.  
Then, the classical Sweet--Parker~\citep{Sweet-1958, Parker-1957} thickness of the reconnection layer is as small as 
$\delta_{\rm SP}=L\,S^{-1/2}\sim 10^{-2}~{\rm cm}$.  
This thickness is  tiny compared with~$L$; however, it is huge compared with all the relevant 
micro-physical plasma length-scales, such as the photon and electron 
collisional mean free paths ($\lambda_{\rm mfp}\sim 10^{-6}~{\rm cm}$),
the electron gyro-radius ($\rho_e\sim 10^{-10}~{\rm cm}$),
and the electron collisionless skin-depth 
($d_e\equiv c/\omega_{pe}\sim 10^{-9}~{\rm cm}$).
This means that, although in absolute terms they are very small, 
the classical collisional resistivity (due to $e^+e^-$ Coulomb 
collisions) and the Compton drag (due to $e^+\gamma$ 
and $e^-\gamma$ collisions) dominate over all other non-ideal 
terms in the generalized Ohm law. Therefore, collisional radiative resistive MHD 
should apply and provide an accurate description of the plasma inside the very thin reconnection layer. 

It is of course, understood that the above argument is not complete. It ignores several important physical ingredients, such as  radiative cooling of the reconnection layer, special-relativistic effects, and the QED effects of the near-critical magnetic field. However, at the very least, the above argument should be considered a cautionary note about the applicability of fast collisionless reconnection theory to this environment.

The main implication of this finding is that reconnection in super-critical fields should proceed in the resistive
plasmoid-dominated regime. Then, provided that our understanding of plasmoid-dominated reconnection can be extended to the highly relativistic and optically-thick electron-positron plasma in an ultra-strong magnetic field (which is a big "if"), one is tempted to conclude that the effective reconnection rate should be bracketed between the purely-resistive rate of~$0.01 \, B_0 V_A$ and the collisionless rate of order~$0.1 \, B_0 V_A$. Which one of these limits applies should depend on the ratio $\delta_c/d_e \sim 100\, \eta V_A/d_e$. One can easily check that, for the physical conditions expected in the environments under consideration, these two scales ($\delta_c$ and~$d_i$) are not very different from each other. Therefore, given the large number of uncertainties in our understanding of reconnection in this regime, it is probably premature to try to draw any definitive conclusions. 

What are the implications of the above range in reconnection rates ($0.01-0.1\, B_0 V_A$) for the main astrophysical systems of interest --- GRBs and magnetar flares? 

In the case of GRB central engines, the high collisionality may suppress immediate very fast large-scale reconnection. Even if the effective reconnection rate is as high as $0.01\, B_0 \, V_A$, a magnetic-tower-like outflow might not be completely destroyed by reconnection and may survive its propagation through the inner region (of order $10^8$~cm) of the star, at which point the outflow becomes relativistic and this results in a slow-down of reconnection in the laboratory frame~\citep{Uzdensky_MacFadyen-2006}. However, as the outflow expands and eventually breaks out from the star, the reconnecting magnetic field decreases and the reconnection layer cools. At some point, pairs annihilate and the particle density drops rapidly. 
Then, the layer becomes collisionless and can switch to a much faster reconnection regime~\citep{Uzdensky_MacFadyen-2006, McKinney_Uzdensky-2010}. The resulting efficient reconnection may transform the outflow into a train of plasmoids~\citep{Uzdensky_MacFadyen-2007b},  accompanied by the corresponding delayed magnetic energy release. Indeed, post-breakout reconnection in relativistic Poynting-flux-dominated jets has been invoked as a plausible mechanism for powering GRB prompt emission~\citep{Lyutikov_Blackman-2001, Lyutikov_Blandford-2002, Lyutikov_Blandford-2003, Drenkhahn_Spruit-2002, Giannios_Spruit-2005, Giannios_Spruit-2006, Giannios_Spruit-2007,  Lyutikov-2006b, McKinney_Uzdensky-2010}.

As for the giant SGR flares, the effective reconnection rate of $cE\sim 0.01\, B_0\, V_A$ translates into a very short characteristic reconnection timescale of about~$3\, {\rm msec}$. This implies that reconnection itself is {\it not} the main energy-release bottleneck during the flare and hence that it does not govern the flare duration. Instead, this duration is probably governed by the escape of radiation  from the hot, relativistically expanding fireball formed as a result of reconnection, and/or by the relatively slow driving by the neutron star crustal motions~\citep{Thompson_Duncan-1995,Thompson_Duncan-2001, Lyutikov-2006a}.

%************************************************************************

\section{Summary}
\label{sec-summary}

The main goal of this paper is to introduce a new frontier in magnetic reconnection research --- 
reconnection in high-energy-density radiative plasmas --- which should be viewed as a fundamental plasma physics problem with important applications to astrophysics. I would like to encourage more research efforts in this area, including numerical simulations and laboratory experiments, and express hope that we will see a rapid progress in the coming years in both our fundamental understanding of it and in astrophysical applications.

In this paper, I first discussed the key physical processes that, although usually not included in traditional studies of magnetic reconnection in heliospheric and laboratory plasmas, become very important and even dominant in some high-energy astrophysical environments. The relative importance of these processes can be linked to the {\it high energy density} of the magnetic field in these systems. These physical processes are: special-relativistic effects, radiation (radiative cooling, radiation pressure, and Compton resistivity), and pair creation.  

I then specifically focused on the case of reconnection involving magnetar-strength magnetic fields, 
i.e., fields exceeding the critical quantum magnetic field $B_* = m_e^2 c^3/e\hbar \simeq 4.4\times 10^{13}\, {\rm G}$.
This case is relevant to several topics in high-energy astrophysics, most notably, to reconnection in magnetar magnetospheres (e.g., SGR giant flares), and to magnetic models of the central engines of GRBs and core-collapse supernovae, as well as the inner parts of magnetically-driven GRB jets. 
In the latter case, understanding reconnection is crucial for determining whether strongly magnetized jets
can survive long enough to break out from the star and for assessing where and how their magnetic energy can eventually be dissipated.

An addition to this strong astrophysical motivation, another reason why I focused on this case is that it represents the most extreme example of high-energy-density astrophysical reconnection, in which all of the above-mentioned physical processes (and also some others!) come into play. Thus, the problem of reconnection of super-critical magnetic fields presents an exciting intellectual challenge with rich and exotic physics. 

The main focus of this paper was on identifying and characterizing quantitatively the basic physical processes 
that are important in this exotic physical environment and on investigating how these processes interact with each other
in the presence of an ultra-strong magnetic field.
In particular, I showed that, because of the extremely high energy density corresponding to a super-critical magnetic field, the main outcome of the dissipation of this energy is the creation of relativistically hot electron-positron-photon plasma in the reconnection layer. The density of photons and pairs in the layer is expected to be so high that the plasma becomes highly collisional and a local thermodynamic equilibrium between mildly degenerate pairs and radiation is easily established inside the layer. Furthermore, large numbers of pairs are also expected to be created by escaping photons even outside (upstream) of the current layer proper, thus effectively dressing the layer in an optically thick "pair coat". Under some conditions, however, when the optical depth is not very high, the radiative cooling by the photons diffusing across the layer may be stronger than the usual advection of energy along the layer. In this strong-cooling regime the problem of the perpendicular structure of the reconnection layer effectively becomes a (one-dimensional) optically-thick radiative transfer problem, coupled with the pressure balance condition. 
Furthermore, even without solving this problem in its entirety, in the case with no guide field and assuming steady state and ignoring QED effects one can establish the following important relationship between the total optical depth of the dressed layer~$\tau$, and the reconnection rate~$E$: $E \sim B_0/\tau$, where $B_0$ is the reconnecting component of the magnetic field. In reality, however, one would have to perform a more thorough analysis taking into account various QED effects on the propagation of high-energy $\gamma$-photons in the presence of a super-critical magnetic field, such as photon splitting and one-photon pair creation. 

In addition, the very high density of pairs and photons inside the layer ensures that the layer is strongly collisional from the reconnection perspective, i.e., $\delta_{\rm SP} \gg d_e, \rho_e$. This implies that radiative resistive MHD should apply within the layer and, therefore, the reconnection process should proceed in the resistive plasmoid-dominated regime, probably at an  expected effective reconnection rate of about~$10^{-2}\,B_0 V_A$. However, because of the large number of uncertainties related to the operation of plasmoid-dominated reconnection in the presence of a strong radiative cooling and of the other processes mentioned above, it is probably premature to make specific claims and predictions. 
Additional processes that need to be taken into account include:  resistivity due to the photon drag, special-relativistic effects, and possibly even neutrino cooling. This is largely an uncharted territory and surprises are expected!

In conclusion, extreme astrophysical reconnection in ultra-strong magnetic fields is a new, physically rich and exciting area of research at the intersection of traditional plasma physics, high-energy astrophysics, and high-energy-density physics. In addition to its astrophysical applications, I hope that this problem will stimulate further studies of magnetic reconnection in the High-Energy-Density context.

%**************************************************************

%**************************************************************

%--------------------------------------------------------------------

%-------------------------------------------------------------------------

%\acknowledgments

%\begin{acknowledgments}

\vskip 15 pt 
I express my gratitude to the organizers of the Yosemite-2010 Workshop on Magnetic Reconnection 
(held on February 8-12, 2010, in Yosemite National Park, California) for inviting me to give a tutorial review talk on which this article is based. I also would like to acknowledge fruitful and encouraging discussions with M.~Barkov, A.~Beloborodov, J.~Drake, J.~McKinney, and M.~Medvedev. This work is supported in part by National Science Foundation Grant  PHY-0903851.

%\end{acknowledgments}

%-------------------------------------------------------------------

%Distribution List: 

%-------------------------------------------------------------------

%\bibliographystyle{apj}aps-bibstyle
%\bibliographystyle{natbib}

%\bibliographystyle{aps-nameyear}      % basic style, author-year citations
%\bibliography{yosemite2}   % name your BibTeX data base

\nocite{*}

%-------------------------------------------------------------------

\end{document}